\documentclass[12pt,preprint,notoc,nohyper]{QCEMDA} 

\usepackage{epsfig,multicol,bbm}

\newcommand\fverb{\setbox\pippobox=\hbox\bgroup\verb}
\newcommand\fverbdo{\egroup\medskip\noindent%
            \fbox{\unhbox\pippobox}\ }
\newcommand\fverbit{\egroup\item[\fbox{\unhbox\pippobox}]}
\newbox\pippobox

\title{Quantum corrections to the entropy of Einstein-Maxwell dilaton-axion black
holes}
 
\author{M. Akbar$^a$ and K. Saifullah$^b$ \\

$^a$Centre for Advanced Mathematics and Physics \\
  National University of Sciences and Technology, Rawalpindi, Pakistan \\
$^b$Department of Mathematics, Quaid-i-Azam University, Islamabad,
Pakistan \\

Electronic address: \email{makbar@camp.nust.edu.pk},
\email{saifullah@qau.edu.pk}}

\preprint{}  

\abstract{We study the corrections to the entropy of
Einstein-Maxwell dilaton-axion black holes beyond semiclassical
approximations. We consider the entropy of the black hole as a state
variable and derive these corrections using the exactness criteria
of the first law of thermodynamics. We note that from this general
frame-work the entropy corrections for ``simpler'' black holes like
Schwarzschild, Reissner-Nordstr\"{o}m and anti-de
Sitter-Schwarzschild black holes follow easily. This procedure gives
us the modified area law as well.}

\begin{document}

\section{Introduction}

In issues like black hole evaporation, Hawking radiation
\cite{Hawking} and quantum tunneling we resort to semiclassical
treatment to study changes in thermodynamical quantities. Thus
quantum corrections to the Hawking temperature and the
Bekenstein-Hawking area law for the Schwarzschild, anti-de Sitter
Schwarzschild, Kerr \cite{Banerjee08} and Kerr-Newman \cite{solod1,
Banerjee09} black holes have been studied in the literature.
Uncharged BTZ black holes have also been studied for these
corrections \cite{solod2, Modak08}. In our earlier work \cite{MA-KS}
we set up a general procedure for studying corrections to the
entropy of charged and rotating black holes beyond semiclassical
approximations. Corresponding modification in the Hawking
temperature and the Bekenstein-Hawking area law were also presented.

Let us write the first law of thermodynamics for charged and
rotating black holes. For the three parameters $M, J, Q$, the mass,
angular momentum and charge of the black hole, respectively, this
can be written as

\begin{equation}\label{2law}
   dM=TdS+\Omega dJ+\Phi dQ ,
\end{equation}
where, $T$ is the temperature, $S$ entropy, $\Omega$ angular
velocity and $\Phi$ electrostatic potential of the black hole. We
can also write this as
\begin{equation}\label{2laws}
   dS(M, J, Q)=\frac{1}{T}dM-\frac{\Omega}{T}dJ-\frac{\Phi}{T}dQ .
\end{equation}
Compare this with the three dimensional differential of a function
$f$,

\begin{equation}\label{diff}
   df(x,y,z)=A(x,y,z)dx+B(x,y,z)dy+C(x,y,z)dz .
\end{equation}
Now, this differential is exact if the following conditions hold

\begin{equation}\label{3cond}
    \frac{\partial A}{\partial y}=\frac{\partial B}{\partial x} ,
\,\,\,\,\,\,\, \frac{\partial A}{\partial z}=\frac{\partial
C}{\partial x} , \,\,\,\,\,\,\,  \frac{\partial B}{\partial
z}=\frac{\partial C}{\partial y} .
\end{equation}
Here we have

\begin{eqnarray}
\frac{\partial f}{\partial x} = A,  \frac{\partial f}{\partial y} =
B,  \frac{\partial f}{\partial z} = C,
\end{eqnarray}
and (\ref{diff}) can be integrated to yield $f$.

If we replace the $A, B, C$ of conditions (\ref{3cond}) by $1/T,
-\Omega/T, -\Phi/T$, in which case $M, J, Q$ will play the role of
$x, y, z$, respectively, we note that in order for $dS$ to be an
exact differential (\ref{2laws}) the following conditions must be
satisfied

\begin{eqnarray}\label{3conda}
    \frac{\partial }{\partial J}\left(\frac{1}{T}\right)
=\frac{\partial }{\partial M} \left(-\frac{\Omega}{T}\right),  \\
\label{3condb} \frac{\partial }{\partial
Q}\left(\frac{1}{T}\right)=\frac{\partial }{\partial M}
\left(-\frac{\Phi}{T}\right),
\\ \label{3condc} \frac{\partial}{\partial
Q}\left(-\frac{\Omega}{T}\right)=\frac{\partial }{\partial
J}\left(-\frac{\Phi}{T}\right) .
\end{eqnarray}
This allows us to write entropy $S(M,J,Q)$ in the integral form. We
employed \cite{MA-KS} this procedure to work out quantum corrections
of entropy beyond the semiclassical limit, of the Kerr-Newman and
the charged rotating BTZ black holes. Further, we showed that the
(quantum) corrections for simpler black holes, found earlier using
different techniques, can be easily recovered as special cases of
that study. In this paper, after briefly describing our earlier
results, we extend this analysis to the axially symmetric
Einstein-Maxwell dilaton-axion black holes \cite{strom, jialing}.
The quantum corrections to the entropy of these black holes have
been investigated using brick wall model \cite{shenchen}. The
correction upto the second term only have been calculated in this
paper. Our method is much simpler and more general at the same time.
We have also presented the modified Bekenstein-Hawking area law. The
leading order correction term is found to be logarithmic \cite{KM}
and the higher order terms have ascending powers of inverse of the
area.

\section{The Kerr-Newman black hole}

The Kerr-Newman spacetime in Boyer-Lindquist coordinates
$(t,r,\theta, \phi)$ can be written as

\begin{eqnarray*}
ds^{2} &=& -\frac{\Delta^2}{\rho^2}(dt-asin^2 \theta
d\phi)^2+\frac{\rho^2}{\Delta^2}dr^2+\rho^2d\theta^{2}+
\frac{sin^2\theta}{\rho^2}(adt-(r^2+a^2)d\phi)^2 ,
\end{eqnarray*}
where

\begin{eqnarray}
 \nonumber \Delta(r)^2 &=& (r^2+a^2)-2Mr+Q^2 , \\
 \nonumber  \rho^2(r,\theta) &=& r^2 + a^2cos^2\theta , \\
 \nonumber  a &=& \frac{J}{M} .
\end{eqnarray}
The inner and outer horizons for this metric are

\begin{equation}\label{eh}
    r_\pm=M\pm\sqrt{M^2-a^2-Q^2} .
\end{equation}
The Hawking temperature is defined as

\begin{equation}\label{tempg}
 T= \left(\frac{\hbar}{4\pi}\right)\left( \frac{r_+-r_-}{r_+^2+a^2}\right) ,
\end{equation}
which, in our case takes the form

\begin{equation}\label{temp}
 T= \left(\frac{\hbar}{2\pi}\right) \frac{\sqrt{M^4-J^2-Q^2 M^2 }}{M\left(2M^2-Q^2+
 2\sqrt{M^4-J^2-Q^2 M^2}\right)} .
\end{equation}
The angular velocity \cite{carroll04}, $\Omega=a/(r_+^2+a^2)$, takes
the form

\begin{equation}\label{omega}
 \Omega= \frac{J}{M\left(2M^2-Q^2+2\sqrt{M^4-J^2-Q^2 M^2 }\right)} ,
\end{equation}
and the electrostatic potential, $\Phi=r_+Q/(r_+^2+a^2)$, becomes

\begin{equation}\label{phi}
 \Phi= \frac{Q \left(M^2+\sqrt{M^4-J^2-Q^2 M^2 }\right)}
{M\left(2M^2-Q^2+2\sqrt{M^4-J^2-Q^2 M^2 }\right)} .
\end{equation}

It is easy to see that these quantities for the Kerr-Newman black
hole satisfy conditions (\ref{3conda})-(\ref{3condc}), and
therefore, $dS$ is an exact differential. Thus we apply the
procedure described in Section 1, and use the corrected form of the
Hawking temperature \cite{Banerjee08}

\begin{equation}\label{int13}
T_c =T \left(1+\sum\frac{\alpha_i
\hbar^i}{\left(r_+^2+a^2\right)^i}\right)^{-1} ,
\end{equation}
where $\alpha_i$ correspond to higher order loop corrections to the
surface gravity of black holes $\mathcal{K} = 2 \pi T$. The modified
surface gravity \cite{York85} due to quantum effects becomes

\begin{equation}\label{sgcor}
  \mathcal{K}=\mathcal{K}_0 \left(1+\sum_i \frac{\alpha_i
\hbar^i}{(r_+^2+a^2)^i}\right)^{-1} .
\end{equation}
Thus the entropy including the correction terms becomes

\begin{eqnarray}\label{soln6}
S &=& \frac{\pi}{\hbar}(r_+^2+a^2)+ \pi \alpha_1 ln (r_+^2+a^2)+
\sum_{k>2} \frac{\pi \alpha_{k-1}
\hbar^{k-2}}{(2-k)(r_+^2+a^2)^{k-2}} + \cdots.
\end{eqnarray}

Note that if we put charge $Q=0$, we recover the corrections for the
case of the Kerr black black hole \cite{Banerjee08}. If the angular
momentum is also zero we get results for the Schwarzschild black
hole ($a=Q=0$). However, if only the angular momentum vanishes (i.e.
$a=0$), we get corresponding corrections for Reissner-Nordstr\"{o}m
black hole, in which case the power series involve the charge $Q$
also, in addition to $M$.

Using the Bekenstein-Hawking area law relating entropy and horizon
area, $S=A/4\hbar$, where the area in our case is

\begin{equation}\label{bh2}
    A=4\pi (r_+^2+a^2) ,
\end{equation}
from (\ref{soln6}) we obtain

\begin{eqnarray}\label{soln7}
S &=& \frac{A}{4 \hbar}+ \pi \alpha_1 ln A -\frac{4 \pi^2 \alpha_2
\hbar}{A}-\frac{8 \pi^3 \alpha_3 \hbar^2}{A^2}-\cdots  ,
\end{eqnarray}
which gives quantum corrections for the area law.

As regards the value of the prefactor $\alpha_i$'s there are
different interpretations found in the literature. For example, some
authors take $\alpha_1$ to be negative \cite{GM}, some positive
integer \cite{Hod}, while others find it to be zero even \cite{Med}.

\section{Charged and rotating BTZ black hole}

The Ba\~{n}ados-Teitelboim-Zanelli (BTZ) black hole \cite{BTZ92},
which is $(1+2)$-dimensional, when it is charged and rotating, can
be written as \cite{BTZchrg}

\begin{eqnarray}
\nonumber
ds^{2}&=&-(-M+\frac{r^2}{l^2}+\frac{J^2}{4r^2}-\frac{\pi}{2}Q^2
\ln{r})dt^{2}\\ &+&
(-M+\frac{r^2}{l^2}+\frac{J^2}{4r^2}-\frac{\pi}{2}Q^2 \ln{r})^{-1}dr
^{2}+r^2(d\phi-\frac{J}{2r^2}dt)^2 ,
\end{eqnarray}
where $M$ is the mass, $J$ the angular momentum, $Q$ the charge of
the black hole, and $1/l^2=-\Lambda$ is the negative cosmological
constant.

The event horizon and the inner horizon $r_+$ and $r_-$ satisfy

\begin{equation}\label{evh}
-M+\frac{r^2}{l^2}+\frac{J^2}{4r^2}-\frac{\pi}{2}Q^2 \ln{r}=0 .
\end{equation}
We write

\begin{equation}
f(r)=-M+\frac{r^2}{l^2}+\frac{J^2}{4r^2}-\frac{\pi}{2}Q^2 \ln{r} .
\end{equation}
The angular velocity of the BTZ black hole is

\begin{equation}\label{avel1}
\Omega= -\left. \frac{g_{\phi t}}{g_{\phi
\phi}}\right|_{r=r_+}=\left. \frac{J}{2 r^2}\right|_{r=r_+} .
\end{equation}
The event horizon is related with temperature $T$ by \cite{KSP,
akbar07}

\begin{eqnarray}\label{btzt}
T=\left. \frac{\hbar f'(r)}{4\pi}\right|_{r=r_+} ,
\end{eqnarray}
where $f'(r)$ denotes the derivative of $f$ with respect to $r$. The
electric potential is given by \cite{BTZchrg}

\begin{equation}\label{ep}
\Phi= -\left. \frac{\partial M}{\partial Q}\right|_{r=r_+}= - \pi Q
\ln r_+ .
\end{equation}

With these thermodynamic quantities the BTZ black hole satisfies the
first law of thermodynamics of the form (\ref{2law}) and the entropy
with quantum corrections is given by the series

\begin{equation}\label{btzs6}
 S= \frac{4 \pi r_+}{\hbar}+4 \pi \alpha_1 \ln r_+-\frac{4 \alpha_2 \hbar
\pi}{r_+}- \cdots .
\end{equation}

Putting $8G_3=1$, where $G_3$ is the three dimensional Newton's
gravitational constant, the area formula is

\begin{eqnarray}
A=2 \pi r_+  .
\end{eqnarray}
If we include $G_3$ this becomes

\begin{eqnarray}
A= 16 \pi G_3 r_+ ,
\end{eqnarray}
and the above result for entropy becomes

\begin{equation}\label{btzs7}
 S= \frac{A}{4 \hbar G_3}+4 \pi \alpha_1 \ln A-\frac{64 \alpha_2 \hbar
\pi^2 G_3}{A}- \cdots .
\end{equation}

It may be pointed out here that these black holes are not physically
significant, however, the above results provide useful mathematical
insights for lower dimensional gravity theories.

\section{Axially symmetric Einstein-Maxwell dilaton-axion black
hole}

The stationary axisymmetric Einstein-Maxwell  black holes in the
presence of dilaton-axion  field are found in heterotic string
theory \cite{strom}. In Boyer-Lindquist coordinates $(t, r, \theta,
\phi)$, these are described by the line element \cite{jialing}

\begin{eqnarray}\nonumber
ds^{2} &=& - \frac{\Sigma - a^{2}sin^{2}\theta}{\Delta} dt^{2} -
\frac{2a sin^{2}\theta}{\Delta}\left[(r^{2}
-2Dr+a^{2})-\Sigma\right]dt d\phi\\ &+& \frac{\Delta}{\Sigma} dr^{2}
+ \Delta d\theta^{2}  + \frac{sin^{2}\theta}{\Delta}
\left[(r^{2}-2Dr+a^{2})^2-\Sigma a^{2}sin^{2}\theta\right]d\phi^{2}
,
\end{eqnarray}
where

\begin{equation}
\Delta = r^{2} -2Dr + a^{2}cos^{2}\theta ,
\end{equation}

\begin{equation}
\Sigma = r^{2}-2Mr + a^{2} .
\end{equation}
They have the electric charge

\begin{equation}
Q = \sqrt{2\omega D(D-M)}, where \omega = e^{d} .
\end{equation}
Here $D$, $M$, $a$ and $d$ denote the dilaton charge, mass, angular
momentum per unit mass and the massless dilaton field, respectively,
and $m = M-D$ is the Arnowitt-Deser-Misner (ADM) mass of the black
hole. The electrostatic potential is

\begin{eqnarray*}
 \Phi = \frac{-2DM}{Q(r_{+}^{2} -2Dr_{+}+ a^{2})} .
\end{eqnarray*}

The metric has singularities at $r^{2} -2Dr+ a^{2}cos^{2}\theta =
0$. The outer and inner horizons are respectively

\begin{equation}\label{horda}
r_{\pm} = \left(M-\frac{Q^2}{2\omega M}\right) \pm
\sqrt{\left(M-\frac{Q^2}{2\omega M}\right)-a^2} .
\end{equation}
The outer horizon at $r_{+}$ is specified as a black hole horizon
and is a null stationary 2-surface. The Killing vector normal to
this surface is $\chi^{\alpha} = t^{\alpha} + \Omega \phi^{\alpha}$
and it is null on the horizon. The angular velocity on the horizon
is given by

\begin{eqnarray*}
 \Omega = \frac{J/M}{ r_{+}^{2} -2Dr_{+}+ a^{2}}
\end{eqnarray*}
or
\begin{eqnarray*}
 \Omega = \frac{J}{2M\left[M(M+D)+\sqrt{M^2(M+D)^2-J^2}\right]}
\end{eqnarray*}

This horizon is generated by the Killing vector $\chi^{\alpha}$, and
the surface gravity $\kappa$ associated with this Killing horizon is
given [15] by

\begin{equation}
\kappa^{2} = \frac{-1}{2} \chi^{\alpha;\beta}\chi_{\alpha;\beta} .
\end{equation}
Using this definition of the surface gravity, it is easy to evaluate
the temperature $T = \kappa/2$ associated with this horizon as,

\begin{equation}
T = \frac{\hbar}{4\pi }\left[\frac{(r_{+}-M-D)}{(r_{+}^{2} -2Dr_{+}+
a^{2})}\right] ,
\end{equation}
or

\begin{equation}\label{tempda}
T = \frac{\hbar}{4\pi }\left[\frac{
\sqrt{M^2(M+D)^2-J^2}}{M[M(M+D)+\sqrt{M^2(M+D)^2-J^2]}}\right] .
\end{equation}

One can easily check that the above thermodynamical quantities
satisfy conditions (\ref{3conda})-(\ref{3condc}). Thus the entropy
differential $dS$ is exact and we can evaluate the integral to work
out the semiclassical entropy

\begin{eqnarray}\label{soln20}
\nonumber
 S(M,J,Q)&=& \int \frac{dM}{T} \\
\nonumber &=& \frac{4\pi}{\hbar}\int
\frac{M\left[M(M+D)+\sqrt{M^2(M+D)^2-J^2}\right]}{\sqrt{M^2(M+D)^2-J^2}} dM \\
&=& \frac{2\pi}{\hbar}\left[M(M+D)+\sqrt{M^2(M+D)^2-J^2}\right] .
\end{eqnarray}

Now, in order to work out the quantum corrections to this formula,
we need to show that $T, \Omega, \Phi$ satisfy the following
conditions involving corrections and which replace conditions
(\ref{3conda})-(\ref{3condc}) for exactness of entropy

\begin{eqnarray*}
\frac{\partial }{\partial J}\frac{1}{T}\left(1+\sum \frac{\beta_i
\hbar^i} {(r_+^2-2Dr_{+}+a^2)^i}\right) =\frac{\partial }{\partial
M} \frac{-\Omega}{T} \left(1+\sum \frac{\beta_i
\hbar^i}{(r_+^2-2Dr_{+}+a^2)^i}\right)
\\ \label{3condKN2} \frac{\partial }{\partial Q}\frac{1}{T}\left(1+\sum \frac{\beta_i
\hbar^i}{(r_+^2-2Dr_{+}+a^2)^i}\right)=\frac{\partial }{\partial M}
\frac{-\Phi}{T}\left(1+\sum \frac{\beta_i
\hbar^i}{(r_+^2-2Dr_{+}+a^2)^i}\right)  \\
\label{3condKN3} \frac{\partial}{\partial
Q}\frac{-\Omega}{T}\left(1+\sum \frac{\beta_i
\hbar^i}{(r_+^2-2Dr_{+}+a^2)^i}\right)=\frac{\partial }{\partial
J}\frac{-\Phi}{T}\left(1+\sum \frac{\beta_i
\hbar^i}{(r_+^2-2Dr_{+}+a^2)^i}\right) .
\end{eqnarray*}

We note that this is indeed the case and the entropy integral is
simplified to
\begin{eqnarray*}\label{soln4}
 S(M,J,Q)&=& \int \frac{1}{T}\left(1+\sum \frac{\beta_i
 \hbar^i}{(r_+^2-2Dr_{+}+a^2)^i}\right)dM ,
\end{eqnarray*}
which can be written in the expanded form as

\begin{eqnarray*}\label{soln50}
 \nonumber S(M,J,Q)&=& \int \frac{1}{T}dM +\int \frac{\beta_1
\hbar}{T(r_+^2-2Dr_{+}+a^2)}dM \\ &+& \int \frac{\beta_2
\hbar^2}{T(r_+^2-2Dr_{+}+a^2)^2}dM \\ \nonumber &+& \int
\frac{\beta_3 \hbar^3}{T(r_+^2-2Dr_{+}+a^2)^3}dM +\cdots \\ &=&
I_1+I_2+I_3+I_4+\cdots ,
\end{eqnarray*}
where the first integral $I_1$ has been evaluated in (\ref{soln20}).
We work out the other integrals one by one after substituting values
from (\ref{horda}) and (\ref{tempda}). Thus

\begin{eqnarray*}\label{int10}
 I_2&=& 2\pi \beta_1 \hbar \int \frac{M dM}{\sqrt{M^2(M+D)^2-J^2}}
\end{eqnarray*}
After making some appropriate substitution this can be evaluated as

\begin{eqnarray*}\label{int110}
I_2&=& \pi \beta_1 \ln \left| M(M+D)+\sqrt{M^2(M+D)^2-J^2} \right|,
\end{eqnarray*}
which is nothing but

\begin{eqnarray}\label{int120}
I_2&=&  \pi \beta_1 \ln (r_+^2-2Dr_{+}+a^2) .
\end{eqnarray}

The k-th integral $I_k$ where $k=3,4,\cdots$ can be evaluated as

\begin{eqnarray*}\label{intk0}
 \nonumber I_k&=&  \int \frac{\beta_{k-1} \hbar^{k-1} dM}{T(r_+^2-2Dr_{+}+a^2)^{k-1}} \\
\nonumber &=& \frac{2\pi}{\hbar}\int \frac{\beta_{k-1} \hbar^{k-1}
M dM}{\sqrt{M^2(M+D)^2-J^2}\left[  M(M+D)+\sqrt{M^2(M+D)^2-J^2}\right]^{k-2}} \\
&=& \frac{\pi \beta_{k-1} \hbar^{k-2}}{2-k} \left[
M(M+D)+\sqrt{M^2(M+D)^2-J^2}\right]^{2-k} , k>2 ,
\end{eqnarray*}
or, in terms of $r_+$, this can be written as

\begin{eqnarray}\label{intk10}
I_k&=&  \frac{\pi \beta_{k-1}
\hbar^{k-2}}{(2-k)(r_+^2-2Dr_{+}+a^2)^{k-2}} , k>2 .
\end{eqnarray}
Thus the entropy including the correction terms becomes

\begin{eqnarray*}\label{soln60}
S &=& \frac{\pi}{\hbar}(r_+^2-2Dr_{+}+a^2)+ \pi \beta_1 \ln
(r_+^2-2Dr_{+}+a^2) \\ &+& \sum_{k>2} \frac{\pi \beta_{k-1}
\hbar^{k-2}}{(2-k)(r_+^2-2Dr_{+}+a^2)^{k-2}} + \cdots .
\end{eqnarray*}

The Bekenstein-Hawking entropy associated with this horizon is one
quarter of the area of the horizon surface. It is important to note
that unlike spherical geometry the horizon surface here is not
simply a 2-sphere. The area of the horizon can be computed from the
2-metric on the horizon and it is given by

\begin{equation}
A = 4\pi (r_{+}^{2} -2Dr_{+} + a^{2}) .
\end{equation}
Therefore the corresponding entropy associated with this horizon is

\begin{eqnarray*}
S = \frac{A}{4\hbar} ,
\end{eqnarray*}
so that the modified area law takes the form
\begin{eqnarray}\label{soln70}
S &=& \frac{A}{4 \hbar}+ \pi \beta_1 \ln A -\frac{4 \pi^2 \beta_2
\hbar}{A}-\frac{8 \pi^3 \beta_3 \hbar^2}{A^2}-\cdots .
\end{eqnarray}

\section{Conclusion}

We write the first law of thermodynamics for black holes with three
parameters, mass, charge and angular momentum. Taking this as a
differential of entropy, we apply the criterion for exactness of
differentials in three variables. This enables us to calculate
quantum corrections in entropy for charged and rotating black holes
beyond the semiclassical terms. We have briefly described results
obtained earlier in the case of Kerr-Newman and charged BTZ black
holes. Our main emphasis is to apply this analysis to axially
symmetric Einstein-Maxwell black holes with dilaton-axion charge.
Our procedure helps in evaluating the integrals involving higher
order corrections of entropy. The first term in the power series is
the semiclassical value and the leading order correction term is
logarithmic. This is consistent with the results in the literature
for different black holes, found by using quantum geometry
techniques, field theoretic methods and the brick wall model. The
higher order terms are in ascending powers of $(r_{+}^{2} -2Dr_{+} +
a^{2})^{-1}$. The modified Bekenstein-Hawking area law has also been
derived. Here we have not required corrections in the angular
momentum and the electrostatic potential.

An important feature of this procedure is that, in a sense, it
unifies the earlier approaches and the results for `simpler'
situations can be recovered very easily. Corrections for the
Kerr-Newman spacetime can be obtained by putting $D=0$. For
non-rotating objects like Reissner-Nordstr\"{o}m black hole the
series is in the powers of $(r_+^2)^{-1}$. If the charge and angular
momentum both are put equal to zero, the corrections for the
Schwarzschild black hole are obtained.


\end{document}